# Quantum tunneling of $V(x) = V_0/|x|^\alpha$ singular potential

A. Zh. Muradyan

Laboratory for research and modelling of quantum phenomena, Yerevan State University, 0025, 1 Alex Manoogian, Yerevan, Republic of Armenia

E-mail: muradyan@ysu.am

**Abstract**

Singularity of the potential function makes quantum tunneling, the case $V(x) = V_0/|x|^\alpha$ of which is the subject of this article, mathematically underdetermined. To circumvent the difficulties it introduced in physics, a potential singularity cutoff is often used, followed by a reverse limit transition, or is a suitable self-adjoint extension of the Hamiltonian along the entire coordinate axis made. However, both of them somehow affect the singular nature of the problem, and so I discuss here how quantum tunneling will behave if the original singular nature of the Schrodinger equation left untouched. To do this, I use the property of the probability density current that the singularities are mutually destroyed in it. It is found that the mildly singular potential with $0 < \alpha < 1$ has a finite, but unusual tunnelling transparency, in particular, a non-zero value at zero energy of the incident particle. The tunneling of 1D Coulomb potential ($\alpha = 1$) exhibits infinitely fast and complete oscillation at the zero energy boundary and a suppression to zero in the high-energy limit. In the more singular region with $\alpha > 1$, the tunneling becomes forbidden, thereby repeating the well-known result of the regularized counterparts.

## 1. Introduction

The principal feature of a singular potential [1-3] is the absence of a singularity point in the domain of determining the potential energy function. At the same time, quantum tunnelling also implies a transition through a point of singularity, and therefore some rules for this transition. To do this, the potential cut-off method is used, or the conditions for matching the wave function and its derivative on both sides of the singular point are introduced. The first one replaces the singular form with a regular one with a cropped singular part, for which the transmission and reflection coefficients are calculated, and then a limit transition is made in the expressions of these coefficients, narrowing the truncation width to zero [4-7]. Another method based on the matching condition uses the requirement that physical quantities be represented by Hermitian operators [8-11]. In this way, asserting the existence of a quantum-mechanical average of the singular potential energy, in [12] the singularity is divided into three classes - mildly singular, singular, and extra singular. In terms of $V(x) = V_0 / |x|^\alpha$, they correspond to the parameter ranges $0 < \alpha < 1$, $1 \leq \alpha < 2$, and $\alpha \geq 2$, respectively. It is noted that for the mildly singular class both solutions are regular and, in principle, can be admitted to the problem of tunneling. For the singular class only one solution is regular, and hence directly acceptable into the problem solving procedure. In the extra singular case, both solutions are divergent.

Much attention was paid to the tunneling of 1D-Coulomb potential. Limiting only to a regular solution automatically leads to the absence of a probabilistic current and, consequently, to the impenetrability of the potential barrier. This result was also derived by the method of limiting smoothing of the potential barrier [12] and later supported in a short presentation [13]. Further, considering the antisymmetric distribution of the potential [14], the irregular solution of the problem was transformed into a regular one by some procedure leading to a finite permeability. In [15] this approach is named as ingenious, but not completely justified. The answer [16] proves that the criticism is related to the symmetric, but not to the antisymmetric form of the potential distribution, which was considered there.

In [17], the tunneling problem of the Coulomb potential is approached from the point of view of analytical continuation of solutions through the singularity point in combination with the method of variation of constants. This allows a complete solution of the 1D scattering problem and obtaining any permeability other than the full one: you only need to choose the appropriate type of self-adjoint expansion. [18] reexamined the permeability problem on base of self-adjoint extensions and explicitly defined that the important Dirichlet boundary condition implies an impenetrable origin. Finally, the approach of [19], by analogy with the approach in [14], cancels the singularity by a bilinear form of the wave function, but yields a vanishing transmission.

Besides the Coulomb, quantum tunnelling has also been studied for the inverse quadratic potential $V(x) = V_0 / |x|^2$. In [20], the transmission coefficient is defined for all self-adjoint extensions of the Hamiltonian with condition $0 < V_0 < 3/4$ and found that tunnelling is possible and occurs unless the self-adjoint extension matrix is diagonal. Proceeding from the family of U(2) inequivalent quantization, the possibility of tunnelling under the same conditions is also stated by [21].

Note, however, that the regularization procedures introduced to reconcile the Schrodinger equation with the standard situation for quantum mechanics, to some extent suppress the original content of the singular problem. For this reason, I am studying here the question of what would be the quantum tunneling of the potential within the framework of the Schrodinger equation, but without any manipulation in order to fully comply with the postulates of quantum mechanics. The approach used proceeds from the property of the current density of probability that the singular terms in it balance each other, and therefore the continuity of the current can be spread over the entire coordinate axis, including the origin. Consideration includes all values of $\alpha > 0$, dividing it into the continuous ranges $0 < \alpha < 1$ (mildly singular), $1 < \alpha < 2$ (intermediate singular), $\alpha > 2$ (extra singular) and the integer values $\alpha = 1$ (1D Coulomb) and $\alpha = 2$ (inverse square). A preliminary discussion of this approach in case of 1D Coulomb potential barrier is made in [22].

## 2. The singular potential transmission

The stationary Schrodinger equation for the singular potential reads

$$\frac{d^2 \psi(z)}{dz^2} + \left( \varepsilon - \frac{u_0}{|z|^\alpha} \right) \psi(z) = 0, \qquad (1)$$

where $z$ is normalized to an arbitrary length $l > 0$, and the energy $\varepsilon$ and the "power" of the potential $u_0$ are normalized to the "recoil" energy $E_{rec} = \hbar^2 / 2ml^2$. I'll look for a solution in the form of

$$\psi(z) = \exp(h(z) \pm i\sqrt{\varepsilon}\, z) \quad (2)$$

with an unknown function $h(z)$, which, according to (1), must satisfy the equation

$$h''(z) + (h'(z))^2 \pm i2\sqrt{\varepsilon}\, h'(z) = u_0 |z|^{-\alpha} \quad (3)$$

and the derivative $h'(z)$ of which must disappear at infinity. Equation (3) for general non-integer values of power $\alpha$ does not have exact solutions expressed in terms of known analytical functions. I'll present approximate solution in which the main singular essence of the equation is preserved.

## 2.1 Quantum tunnelling of a mildly singular potential

The leading-order contributions on the right-hand side and the left-hand side of equation (3) near the singular point must be the same order of magnitude, so we have

$$h''(z) \stackrel{z \to 0}{\approx} u_0 |z|^{-\alpha}, \quad (4)$$

and correspondingly

$$h'(z) \stackrel{z \to 0}{\approx} \frac{u_0}{1-\alpha} |z|^{1-\alpha} \quad (5a)$$

and

$$(h'(z))^2 \stackrel{z \to 0}{\approx} \frac{u_0^2}{(1-\alpha)^2} |z|^{2-2\alpha}. \quad (5b)$$

Since $0 < \alpha < 1$, then

$$(h'(z))^2 \ll h'(z) \quad (6)$$

and can be omitted in the equation (3). Then the approximate equation reads

$$h''(z) \pm i2\sqrt{\varepsilon}\, h'(z) = u_0 |z|^{-\alpha}. \quad (7)$$

It should be noted that although condition (6) is gradually violated when moving away from the origin, but since the tunneling problem will be formulated at the origin, it is only important that at long distances $h'(z) \stackrel{z \to \infty}{\sim} |z|^{-\alpha}$ and, accordingly, the solution of equation (7) satisfies the necessary condition of vanishing at infinity.

Equation (7) is of the first order for the derivative $h'(z)$, and its solution for $z > 0$ coordinates is

$$h'(z) = -e^{\mp i 2\sqrt{\varepsilon} z} z^{1-\alpha} (\mp i 2\sqrt{\varepsilon} z)^{-1+\alpha} \Gamma(1-\alpha, \mp i 2\sqrt{\varepsilon} z) u_0, \quad (8)$$

where the constant of integration is taken zero to provide a zero solution in the absence of a tunnelling potential $u_0 = 0$. Then the function $h(z)$ does present as

$$h(z) = -\frac{2^{-2+\alpha} (\mp i \sqrt{\varepsilon})^{\alpha}}{\varepsilon} \left( \frac{\pi \mathrm{Csc}(\pi\alpha)}{\Gamma(\alpha)} \pm \frac{i e^{\mp i 2\sqrt{\varepsilon} z} \Gamma(2-\alpha, \mp i 2\sqrt{\varepsilon} z)}{1-\alpha} \right) u_0. \quad (9)$$

On the left side of the potential, one has

$$h'(z) = \mp \frac{i}{2} e^{\mp i 2\sqrt{\varepsilon} z} (\pm 2i)^{\alpha} \varepsilon^{(-1+\alpha)/2} \Gamma(1-\alpha, \mp i 2\sqrt{\varepsilon} z) u_0, \quad (10)$$

and

$$h(z) = -(\pm i)^{1+\alpha} (2\sqrt{\varepsilon})^{-1+\alpha} \left( \frac{2^{-\alpha} z (\mp i\sqrt{\varepsilon} z)^{-\alpha}}{-1+\alpha} \pm \frac{i(\Gamma(1-\alpha)) - e^{\mp i 2\sqrt{\varepsilon} z} \Gamma(1-\alpha, \mp i 2\sqrt{\varepsilon} z)}{2\sqrt{\varepsilon}} \right) u_0. \quad (11)$$

The linearly independent solutions (2) and their derivatives are finite at the singular point. Therefore, the familiar continuity conditions may be applicable to the general solution of the problem under consideration. Then, assuming the asymptotic absence of a right-propagating wave on the right side of the barrier, the following explicit expressions are obtained for normalized amplitudes of the transmitted and reflected waves:

$$t = \frac{2\varepsilon - (2i)^{\alpha} \varepsilon^{\alpha/2} \Gamma(1-\alpha) u_0}{2\left(2\varepsilon - (-2i)^{\alpha} \varepsilon^{\alpha/2} \Gamma(1-\alpha) u_0\right)} + \frac{1}{2}, \quad (12a)$$

$$r = \frac{2\varepsilon - (2i)^{\alpha} \varepsilon^{\alpha/2} \Gamma(1-\alpha) u_0}{2\left(2\varepsilon - (-2i)^{\alpha} \varepsilon^{\alpha/2} \Gamma(1-\alpha) u_0\right)} - \frac{1}{2}. \quad (12b)$$

Figure 1 shows graphs of the transmission coefficient $T = |t|^2$ and reflection coefficient $R = |r|^2$ as a function of the energy of the incident particle for a potential barrier ($u_0 > 0$). The case of a potential well ($u_0 < 0$) is shown in figure 2.

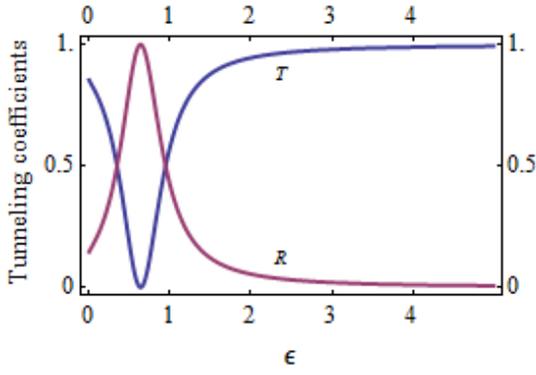

**Figure 1.** Transmission and reflection coefficients of a mildly singular potential barrier as a function of the energy of incident particles for $u_0 = 1$, $\alpha = 0.25$. Appearance of total reflection and non-zero transmission at zero energy is due to the singular nature of the potential function.

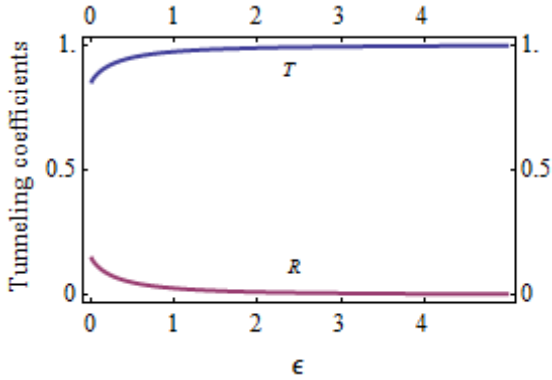

**Figure 2.** Transmission and reflection coefficients of a mildly singular potential well as a function of the energy of incident particles for $u_0 = -1$. The unusual feature here is the non-zero transparency in the zero energy limit. The other parameter $\alpha$ is equal to that used in figure. 1.

Quantum tunneling exhibits obviously unusual behaviour in the low- and moderate-energy ranges: The transmission begins with a non-zero value, and in the case of a potential barrier, the reflection coefficient for a some incident energy increases up to one, which is by no means characteristic of a separate regular potential barrier [23, 24]. The only natural matter here is the ascent to full transparency at asymptotically high energies.

The dependence on the power $u_0$ is also unnatural. Figure 3 illustrates this for both the attractive and repulsive potentials.

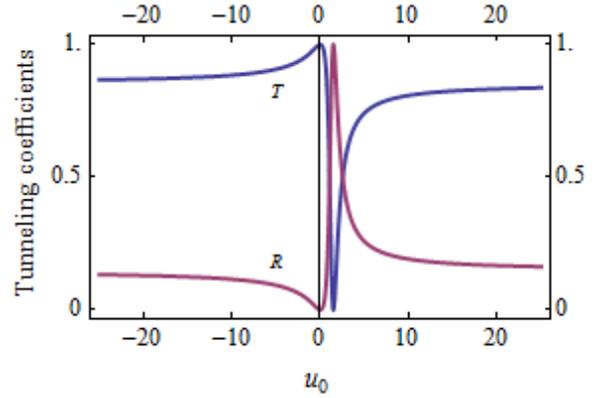

**Figure 3.** The quantum tunneling coefficients of a mildly singular potential depending on the power $u_0$ at $\varepsilon = 1$.

## 2.2 Quantum tunnelling of 1D Coulomb potential

The corresponding stationary state Schrodinger equation reads

$$\frac{d^2 \psi(z)}{dz^2} + \left(\varepsilon - \frac{u_0}{|z|}\right) \psi(z) = 0. \quad (13)$$

Linearly independent solutions of equation (13) on the right side of the origin ($z > 0$) are given by continuously differentiable expressions

$$\psi_{r,1}(z) = e^{-i\sqrt{\varepsilon}z} z\, {}_1F_1\left(1 - \frac{iu_0}{2\sqrt{\varepsilon}}, 2, i2\sqrt{\varepsilon}z\right), \quad (14a)$$

$$\psi_{r,2}(z) = e^{-i\sqrt{\varepsilon}z} z\, U\left(1 - \frac{iu_0}{2\sqrt{\varepsilon}}, 2, i2\sqrt{\varepsilon}z\right), \quad (14b)$$

where $_1F_1(\cdots)$ and $U(\cdots)$ are the hypergeometric functions of Kummer and Tricomi, respectively. The first solution behaves like a regular one: the function itself and its derivative have a finite limit when approaching a singular point. The singular content of the problem manifests itself explicitly in the second solution in the form of logarithmic divergence of its derivative. The general solution is

$$\psi_r(z) = a_{r,1}\psi_{r,1}(z) + a_{r,2}\psi_{r,2}(z), \qquad (15)$$

Left-sided solutions of the equation (13) are obtained by substituting $z \to -z$, i.e.,

$$\psi_{l,1}(z) = -e^{i\sqrt{\varepsilon}z} z\, _1F_1\left(1 - \frac{iu_0}{2\sqrt{\varepsilon}}, 2, -i2\sqrt{\varepsilon}z\right), \qquad (16a)$$

$$\psi_{l,2}(z) = -e^{i\sqrt{\varepsilon}z} z\, U\left(1 - \frac{iu_0}{2\sqrt{\varepsilon}}, 2, -i2\sqrt{\varepsilon}z\right), \qquad (16b)$$

and correspondingly

$$\psi_l(z) = a_{l,1}\psi_{l,1}(z) + a_{l,2}\psi_{l,2}(z). \qquad (17)$$

Since the basis solutions are finite at $z \to \pm 0$, the general solutions (15) and (17) can be considered as continuing each other, i.e. $\psi_l(-0) = \psi_r(+0)$. Given also that $\psi_{l,1}(-0) = \psi_{r,1}(+0) = 0$ and $\psi_{l,2}(-0) = \psi_{r,2}(+0)$, we come to the following relation between the coefficients sought:

$$a_{l,2} = a_{r,2}. \qquad (18)$$

In the following discussion, we leave aside the question of matching the derivative of the wave function and instead accept the continuity condition for the probability density flow

$$j = i\kappa \int (\psi {\psi^*}' - \psi^* \psi') dz, \qquad (19)$$

where $\kappa = \hbar/2ml$ is a constant and the derivative is with respect to the dimensionless coordinate $z$. Direct substitution (15) and (17) in (19) for the left-hand and right-hand current gives

$$j_l = |a_{l,1}|^2 j_{l,11} + a_{l,2}^* a_{l,1} j_{l,12}$$
$$+ a_{l,1}^* a_{l,2} j_{l,21} + |a_{l,2}|^2 j_{l,22}, \qquad (20a)$$

$$j_r = |a_{r,1}|^2 j_{r,11} + a_{r,2}^* a_{r,1} j_{r,12}$$
$$+ a_{r,1}^* a_{r,2} j_{r,21} + |a_{r,2}|^2 j_{r,22}, \qquad (20b)$$

where

$$j_{l(r),mn} = i\kappa \int (\psi_{l(r),m} {\psi_{l(r),n}'}^* - \psi_{l(r),m}' \psi_{l(r),n}^*) dz \qquad (21)$$

is a component of the probability current compiled of the basis states, $m,n = 1,2$.

Since $\psi_{l,1}(z)$ and $\psi_{r,1}(z)$ are real functions, $j_{l,11}$ and $j_{r,11}$ in (20a) and (20b), respectively, are identically zero. Besides, $j_{l,21} = j_{l,12}^*$, $j_{r,21} = j_{r,12}^*$ by definition, and $j_{l,12} = -j_{r,12}$, $j_{l,22} = -j_{r,22}$. In addition, one of the $a$-coefficients in (20a) and (20b) can always be assumed to be known and real. As such, we choose the coefficient $a_{r,2}$. Then the condition of continuity of the probability current looks as

$$j_{l,12} a_{l,1} + j_{l,12}^* a_{l,1}^* = -j_{l,12} a_{r,1} - j_{l,12}^* a_{r,1}^* - 2j_{r,22} a_{r,2} \qquad (22)$$

which is regarded as an interrelation between unknown coefficients $a_{l,1}$ and $a_{r,1}$.

To move forward in determining these coefficients, we turn to the tunneling with a left-sided incident matter wave. The far-distance asymptote $z \to +\infty$ gives

$$\psi_r(z) \approx \frac{2^{-1+\frac{iu_0}{2\sqrt{\varepsilon}}} (i\sqrt{\varepsilon})^{-1-\frac{iu_0}{2\sqrt{\varepsilon}}}}{\Gamma\left(1 - \frac{iu_0}{2\sqrt{\varepsilon}}\right)} a_{r,1} z^{-\frac{iu_0}{2\sqrt{\varepsilon}}} e^{i\sqrt{\varepsilon}z} +$$

$$2^{-1+\frac{iu_0}{2\sqrt{\varepsilon}}} \left(\frac{(-i\sqrt{\varepsilon})^{-1+\frac{iu_0}{2\sqrt{\varepsilon}}}}{\Gamma\left(1 + \frac{iu_0}{2\sqrt{\varepsilon}}\right)} a_{r,1} + (i\sqrt{\varepsilon})^{-1+\frac{iu_0}{2\sqrt{\varepsilon}}} a_{r,2}\right) z^{\frac{iu_0}{2\sqrt{\varepsilon}}} e^{-i\sqrt{\varepsilon}z},$$

where $\Gamma(\cdot)$ is the gamma function. Then the absence of a wave falling from the right side means the zero value of the second line of this expression. It explicitly expresses the unknown coefficient $a_{r,1}$ through the chosen as known $a_{r,2}$, namely

$$a_{r,1} = \frac{(i\sqrt{\varepsilon})^{-1+\frac{iu_0}{2\sqrt{\varepsilon}}}}{(-i\sqrt{\varepsilon})^{-1+\frac{iu_0}{2\sqrt{\varepsilon}}}} \Gamma\left(1+\frac{iu_0}{2\sqrt{\varepsilon}}\right) a_{r,2} \quad . \quad (23)$$

The factor before the amplitude $a_{r,2}$ in (21) is a real quantity, and so is the amplitude $a_{r,1}$. Further, since the amplitude $a_{l,1}$ belongs to the regular part of the general solution (17), to it can be assigned a real character without any influence on the singularity of the solution. Then equation (20) reads as

$$a_{l,1} = -a_{r,1} - \frac{j_{r,22}}{\mathrm{Re}[j_{l,12}]} a_{r,2}, \quad (24)$$

which, given (22), uniquely determines $a_{l,1}$ through the $a_{r,2}$. As a result, equations (18), (23), and (24) define all three amplitudes in terms of $a_{r,2}$ and thus uniquely formulate the quantum tunnelling problem. It is of principal important that the $a$-amplitudes were obtained without touching the question of the derivative of the wave function at the singularity point, and thus without altering singular nature of the problem. As a result, far from the singular point, the incident, transmitted, and reflected parts of the wave function have the following explicit expressions:

$$\psi_{incid} \approx -2^{-1+\frac{iu_0}{2\sqrt{\varepsilon}}} \left( \frac{(i\sqrt{\varepsilon})^{-1+\frac{iu_0}{2\sqrt{\varepsilon}}}}{\Gamma\left(1+\frac{iu_0}{2\sqrt{\varepsilon}}\right)} a_{l,1} \right.$$

$$\left. + e^{2i\pi\left(-1+\frac{iu_0}{2\sqrt{\varepsilon}}\right)} (-i\sqrt{\varepsilon})^{-1+\frac{iu_0}{2\sqrt{\varepsilon}}} a_{l,2} \right) \left(\frac{1}{z}\right)^{-\frac{iu_0}{2\sqrt{\varepsilon}}} e^{i\sqrt{\varepsilon}z},$$

$$\psi_{trans} \approx \frac{2^{-1-\frac{iu_0}{2\sqrt{\varepsilon}}} (i\sqrt{\varepsilon})^{-1-\frac{iu_0}{2\sqrt{\varepsilon}}}}{\Gamma\left(1-\frac{iu_0}{2\sqrt{\varepsilon}}\right)} a_{r,1} z^{-\frac{iu_0}{2\sqrt{\varepsilon}}} e^{i\sqrt{\varepsilon}z},$$

$$\psi_{refl} \approx -\frac{2^{-1-\frac{iu_0}{2\sqrt{\varepsilon}}} e^{\frac{\pi u_0}{\sqrt{\varepsilon}}} (-i\sqrt{\varepsilon})^{-1-\frac{iu_0}{2\sqrt{\varepsilon}}}}{\Gamma\left(1-\frac{iu_0}{2\sqrt{\varepsilon}}\right)} a_{l,1} \left(\frac{1}{z}\right)^{\frac{iu_0}{2\sqrt{\varepsilon}}} e^{-i\sqrt{\varepsilon}z}.$$

They, according to equation (20), determine the desired probabilistic currents and, ultimately, the transmission and reflection coefficients:

$$T = \frac{j_{trans}}{j_{incid}}, \quad R = \frac{-j_{refl}}{j_{incid}} = 1 - T. \quad (25)$$

The tunnelling efficiency shown in figure 4, completely oscillates between zero and one, and the rate of oscillation increases to infinity as it approaches the boundary value $\varepsilon = 0$. And at the high-energy limit, the transmission (reflection) monotonically disappears (becomes complete) [22]. This is how the singularity is reflected in the problem of quantum tunnelling of a one-dimensional Coulomb potential.

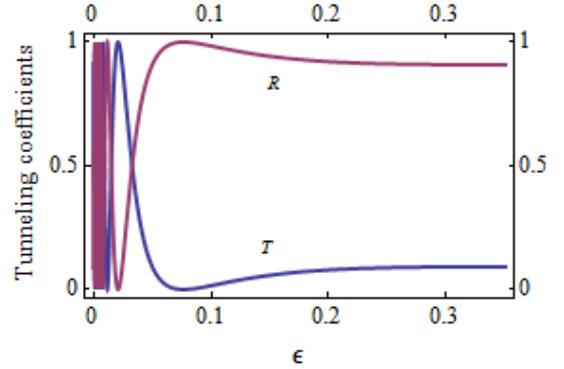

**Figure 4.** Quantum tunnelling coefficients $T$ and $R$ for the 1D Coulomb potential as a function of the energy of the incident particle for both repulsive and attractive cases at $|u_0| = 1$. Unusual features are infinitely accelerating oscillation in the zero energy limit and the approximation of $T$ ($R$) to zero (one) in the high-energy limit. Decreasing $|u_0|$ slows down the oscillation rate.

## 2.3 On the quantum tunnelling of an intermediate singular potential

Near the origin, the asymptotic formulas (4) and (5a) remain valid, from which it follows

$$h(z) \approx \frac{u_0}{(1-\alpha)(2-\alpha)} |z|^{2-\alpha} \underset{z \to 0}{\to} 0. \qquad (26)$$

This, according to equation (2), gives the boundary condition $\psi(z) \underset{z \to 0}{\to} 1$, meaning that in the limit $z \to 0$ both basic solutions are finite. Then, without loss of generality, we can assume the continuity of the general wave function, which is written as the following condition:

$$a_{l,plus} + a_{l,minus} = a_{r,plus} + a_{r,minus}, \qquad (27)$$

where $a_{l(r),plus}$ and $a_{l(r),minus}$ are the probability amplitudes before the basic solutions $\psi_{l(r),plus}(z)$ and $\psi_{l(r),minus}(z)$ of equation (7) to the left and right of the singular point of the potential, respectively. From the absence of a right-hand propagating wave in the asymptotic $z \to +\infty$, the condition $a_{r,minus} = 0$ follows, and then the continuity condition (27) takes a simpler form:

$$a_{l,plus} + a_{l,minus} = a_{r,plus}. \qquad (28)$$

At a long distance, the probability current density is reduced to the expression

$$i\kappa \left( |a_{l,minus}|^2 - |a_{l,plus}|^2 \right) \left( \psi_{l,plus}^* \psi_{l,plus}' - \psi_{l,plus}'^* \psi_{l,plus} \right)$$

to the left of the singular point and to the expression

$$i\kappa |a_{r,plus}|^2 \left( -\psi_{r,plus}^* \psi_{r,plus}' + \psi_{r,plus}'^* \psi_{r,plus} \right)$$

to the right of the singular point. The first expression also took into account that $\psi_{l,minus} = \psi_{l,plus}^*$ and $\psi_{l,plus}' = \psi_{l,plus}'^*$, which directly follows from their explicit expressions. The combinations of wave functions in parentheses differ from zero and are equal with the reverse sign in the general case. Therefore, the equality of probability currents implies

$$|a_{l,plus}|^2 - |a_{l,minus}|^2 = |a_{r,plus}|^2. \qquad (29)$$

The resultant equations (28) and (29) are compatible only if $a_{l,plus} = -a_{l,minus}$. Then equation (28) states that $a_{r,plus} = 0$ -i.e., the intermediate singular potential is completely impenetrable.

## 2.4 On the quantum tunnelling of inverse square and extra singular potentials.

The state equation (1) for the inverse square potential has exact analytical solution:

$$\psi_{r,1}(z) = \sqrt{z} J_\nu \left( \sqrt{\varepsilon} z \right), \qquad (30a)$$

$$\psi_{r,2}(z) = \sqrt{z} Y_\nu \left( \sqrt{\varepsilon} z \right) \qquad (30b)$$

for $z > 0$ and similarly for $z < 0$. Here $J_\nu(\cdot)$ and $Y_\nu(\cdot)$ are Bessel functions of the first and second kind respectively, $\nu = \sqrt{0.25 + u_0}$.

A systematic analysis of solutions shows that for this form of potential function, the solution to the tunnelling problem depends on the sign of $u_0$, that is, on whether the potential is a barrier or a well.

In the case $u_0 > 0$, the solution (30a) is zeroed at the singular point z=0, the solution (30b) tends to infinity, and as a result, the direct application of the continuity condition for the wave function becomes problematic. Here we approach the wave function continuity as a postulate of quantum mechanics, concretizing its meaning in the fact that with an asymptotic approach to a singular point, the corresponding wave functions on the right and left sides would diverge equally. Let's call it the condition of quasi-continuity. In this context, it is important that the quasi-continuity completely preserves the singular content of the wave function and passes into the familiar continuity condition if the problem is regularized. After some math, the quasi-continuity condition reads

$$a_{l,2} = a_{r,2}. \quad (31)$$

The absence of a right-hand propagating wave in the asymptotic $z \to +\infty$ gives the following relationship:

$$a_{r,1} = -i a_{r,2}. \quad (32)$$

And finally, the union of (31) and (32) with the condition of continuity of the probability current gives a summarized statement that

$$\text{Im}[a_{l,1}] = a_{r,2}, \quad \text{Re}[a_{l,1}] \text{ is arbitrary.} \quad (33)$$

A free assumption about the real nature of the coefficient $a_{l,1}$ directly yields $a_{r,1} = a_{r,2} = 0$, that is, the absence of a matter wave on the right side of the singularity point and, accordingly, to the expected result: complete impenetrability of the inverse square singular potential barrier.

In the case of negative $u_0$, both solutions (30a) and (30b) are equal to zero at the singular point $z = 0$. This ultimately makes it impossible to arrive at a definite answer about the possibility of tunnelling an inverse square potential well within the framework of the approach presented in this article.

The extra singular potential with $\alpha > 2$ does not have an exact analytical solution, and we proceed from the approach developed in paragraphs 2.1 and 2.3. The answer essentially repeats the conclusions for the inverse square potential that the extra singular potential barrier is completely impenetrable, and the potential well is beyond the possibility of the approach.

## 3. Conclusion

The problem of quantum tunnelling of a singular potential usually applies the regularization method, when the singularity is first removed in a narrow area around a singular point, the problem is solved for this prototype, and then in the coefficients of transmission and reflection, the limiting transition of narrowing the regularization region to zero is made. The other approach implies physically perceived conditions (Friedrichs extension in the Hilbert space) for matching the wave function and its derivative on both sides of the singularity point. In particular, the one-dimensional Coulomb potential in both approaches turns out to be impenetrable. The preservation of the mathematical essence of the singularity, which is performed in this article, dramatically changes the picture of tunneling in case of mildly singular and one-dimensional Coulomb potentials. In the case of mildly singularity, for example, the transparency of the potential remains finite even at the zero energy boundary. In the other, Coulomb case, the transparency of the potential at the same energy boundary varies infinitely often between one and zero, and with the transition to high energies, it gradually decreases to zero.

And finally, potentials with a higher, $\alpha > 1$ degree of singularity, as in the regularization methods, exhibit complete impenetrability (the case of a potential well at $\alpha \geq 2$ remains outside the framework of the developed approach).

## Acknowledgements

This work was supported by Armenian State Committee of Science. I would like to thank Dr. Gevorg Muradyan for a close discussion.
.

## References

[1] K. M. Case, Singular potentials, Phys. Rev. **80**, 797 (1950).
[2] K. Meetz, Singular potentials in nonrelativistic quantum mechanics," Nuovo Cimento, **34**, 5738 (1964).
[3] W. M. Frank, D. J. Land and R .M. Spector, Singular potentials, Rev. Mod. Phys. **43**, 36 (1971).

[4] R. Loudon, One-dimensional hydrogen atom, Am. J. Phys. **27**, 649 (1959).
[5] L. K. Haines and D. H. Roberts, One-dimensional hydrogen atom, Am. J. Phys. **37**, 1145 (1969).
[6] R. Loudon, One-dimensional hydrogen atom, Proc. Roy. Soc. A, **472**, 20150534 (2016).
[7] G. A. Muradyan, Regularization of quantum tunneling of singular potential barrier, J. Cont. Phys. (Armrnian Academy of Sciences), **54**, 333 (2019).
[8] W. Fisher, H. Laschke, and P. Muller, The functional-analytic versus the functional integral approach to quantum Hamiltonians: The one-dimensional hydrogen atom, J. Math. Phys. **36**, 2313 (1995).
[9] D. M. Gitman, I. V. Tyutin, and B. L. Voronov, *Self-adjoint extensions in quantum mechanics*, 1st ed. (Birkhouser, 2012).
[10] B. Simon, Schrödinger Operators in the twentieth century, J. Math. Phys. **41**, 3523 (2000).
[11] I. Tsutsui, T. Fulop, and T. Cheon. Connection conditions and the spectral family under singular potentials, J. Phys. A: Math. Gen,.**36**, 275 (2003).
[12] M. V. Andrews, Singular potentials in one dimension, Am. J. Phys. **44,** 1064 (1976).
[13] C. L. Hammer and T. A. Weber, Comments on the one-dimensional hydrogen atom, Am. J. Phys. **56,** 281 (1988).
[14] M. Moshinsky, Penetrability of a one-dimensional Coulomb potential, J. Phys. A: Math. Gen. **26** 2445 (1993).
[15] R. G. Newton, Comment on "Penetrability of a one-dimensional Coulomb potential" by M. Moshinsky, J. Phys. A: Math. Gen. **27,** 4717 (1994).
[16] M. Moshinsky, Response to "Comment on 'Penetrability of a one-dimensional Coulomb potential' " by Roger G. Newton, J. Phys. A: Math. Gen..**27,** 4719 (1994).
[17] V. S. Mineev, The physics of self-adjoint extensions: One-dimensional scattering problem for the Coulomb potential, Theor. Math. Phys. **140**, 1157 (2004).
[18] C. R. de Oliveira and A. A. Verri, Self-adjoint extensions of Coulomb systems in 1, 2 and 3 dimensions, Annal. Phys. **324**, 251 (2009).
[19] G. Abramovici and Y. Avishai, The one-dimensional Coulomb problem, J. Phys. A: Math. Gen. **42,** 285302 (2009).
[20] J. Dittrich and P. Exner, Tunneling through a singular potential barrier, J. Math. Phys. **26**, 2000 (1985).
[21] H. Miyazaki and I. Tsutsui, Quantum tunneling and caustics under inverse square potential, Annal. Phys. **299,** 78 (2002).
[22] A. Muradyan and G. Muradyan, On the quantum tunneling of one dimensional Coulomb singular potential barrier, arXiv:2008.12957[quant-ph] (2020).
[23] D. Bohm, *Quantum Theory*, (Prentice-Hall, Engelwood Cliffs,N.Y., 1951), p. 283.
[24] E. Merzbacher, *Quantum Mechanics*, (John Willey & Sons, New York, 1970), p. 100.